\documentclass{article}
\usepackage[T1]{fontenc}
\usepackage[utf8]{inputenc}
\usepackage[italian,english]{babel}
\usepackage{amsmath}
\usepackage{amsthm}
\usepackage{amssymb}
\usepackage{comment}
\usepackage{eucal}
\usepackage{graphicx}
\usepackage{enumerate}
\usepackage{pictex, dcpic}
\usepackage[babel]{csquotes}

\usepackage{color}

\def\al{\alpha}
\def\be{\beta}
\def\de{\delta}
\def\ga{\gamma}

\def\ep{\epsilon}

\def\te{\theta}
\def\la{\lambda}
\def\ze{\zeta}
\def\om{\omega}
\def\si{\sigma}

\def\Ga{\Gamma}

\def\Dal{\Box}

 \def\calU{{\mathcal{U}}}
 
 \def\calC{{\mathcal{C}}}
 \def\calP{{\mathcal{P}}}

 \def\R{{{\mathbb R}}}

 \def\R{{{\mathbb R}}}

\def\Aut{{\hbox{Aut}}}

\def\GL{{\hbox{GL}}}

\def\Diff{{\hbox{Diff}}}

\def\ip{\hbox to4pt{\leaders\hrule height0.3pt\hfill}\vbox to8pt{\leaders\vrule width0.3pt\vfill}\kern 2pt}
 
\def\del{\partial}
\def\na{\nabla}

\def\Vec{\mathfrak{X}}

\def\arr{\rightarrow}

\def\ffrac[#1/#2]{\hbox{$\frac{#1}{#2}$}}
\def\Frac[#1/#2]{\frac{#1}{#2}}
\def\({\left(}
\def\){\right)}
\def\[{\left[}
\def\]{\right]}
\def\^#1{{}^{#1}_{\>\cdot}}
\def\_#1{{}_{#1}^{\>\cdot}}
\def\<{\kern -1pt}

\def\beq{\begin{equation}}
\def\eeq{\end{equation}}
{\left\lbrace\begin{array}{@{}l@{}}}%
{\end{array}\right.}

\begin{document}

\title{Generally Covariant vs. Gauge Structure for Conformal Field Theories}
\author{M.Campigotto$^{a,b}$, L.Fatibene$^{c,b}$
\\
{\small $^a$ Dipartimento di Fisica, University of Torino, Via P. Giuria 1, 10125, Torino, Italy}\\
{\small $^b$ Istituto Nazionale di Fisica Nucleare (INFN), Via P. Giuria 1, 10125, Torino, Italy}\\
{\small $^c$ Dipartimento di Matematica, University of Torino, Via C. Alberto 10, 10123, Torino, Italy}\\
}

\maketitle

\begin{abstract}
We introduce the natural lift of spacetime diffeomorphisms for conformal gravity and 
discuss the physical equivalence between the natural and gauge natural structure of the theory. 
Accordingly, we argue that conformal transformations must be introduced as gauge transformations (affecting fields but not spacetime point)
and then discuss special structures implied by the splitting of the conformal group. 
\end{abstract}

\section{Introduction}

{\it Conformal transformations} are used in two different meanings in the literature.
In some cases, conformal transformations are angle preserving  maps on (pseudo)-Riemannian manifolds. 
In this case they are a subgroup of the manifold diffemorphisms. This viewpoint is common in conformal field theories.
In other cases they are defined as transformations acting on the metric field, which change the metric by multiplying it by a positive (pointwise) factor, called the {\it conformal factor}, but leaving the point on the manifold unchanged. In this case they are vertical maps on the bundles of metrics and sometimes they are called {\it Weyl transformations} to distinguish them from the conformal transformations considered above. This viewpoint is common in conformal gravity. 

If one takes a Minkowskian field theory and tries to enlarge the symmetry group from isometries to a bigger group including angle preserving 
transformations, then the first viewpoint may be meaningful. However, if one is considering a generally covariant theory and aims to a conformal extensions the first attitude does not make sense. The theory is already covariant with respect to general diffeomorphisms, including any angle preserving map.
Thus in this case the second attitude is needed.

Recently, the gauge nature of conformal transformations has been questioned based on the peculiar behaviour of conformal transformations with respect to Noether theorem; see \cite{Jackiw}. 
While this discussion is certainly of interest we shall here consider the issue from a different, complementary perspective, by considering the mathematical consequences of having Weyl transformations acting as symmetries focusing in particular on global properties and trying to provide a rigorous common framework for comparing the two viewpoints presented above. We hope the two different approaches can contribute to setting the issue once and for all.

Mannheim proposed a  generally covariant, conformal  theory for gravity; see \cite{Mannheim}. 
Recently we set this conformal theory, following the second  approach, in a gauge natural form; see \cite{Marta}.
Presenting a theory within the context of gauge natural framework gives some advantages since one has a number of general theorems and standard algorithmic procedures  to compute things and discuss observable quantities. For example, one has a general result claiming that all Noether currents of a gauge natural theory are not only closed on-shell (i.e.~along solutions of field equations) but also exact on-shell so that one has existence in general of the so-called superpotentials; see \cite{Book}, \cite{200Lagrange}, \cite{Augmented}.

The aim of this paper is to discuss the relation between those two viewpoints. In particular we shall show that conformal gravity, beside being a gauge natural theory (see \cite{Book}), also has a natural structure, which defines an action of diffeomorphisms on fields.
Namely, unlike in general gauge natural field theories, in conformal gravity spacetime diffeomorphisms do act on fields  as symmetries. In other words, conformal gravity is somewhere in between a natural and a gauge natural theory.
In particular, one can define by it the action of angle preserving maps as in the first viewpoint and split conformal transformations as the action of a diffeomorphism and a Weyl transformation. 


We shall also discuss whether there is a physical difference between the natural and gauge natural framework and argue that the correct way of introducing conformal gravity is giving it a gauge natural structure as we did, while the natural structure is to be considered as an extra feature of the theory.

In gauge natural theories the symmetry group contains {\it pure gauge transformations} (which do not change position in spacetime, i.e.~they act on fields alone and they are vertical transformations on the configuration bundle). {\it Generalized gauge transformations} instead project on spacetime diffeomorphisms. 
Except from this projection, gauge transformations are disconnected from spacetime diffeomorphisms. In general one cannot even say how spacetime diffeomorphisms act on fields exactly because the symmetry group {\it projects} onto diffeomorphisms instead of having diffeomorphisms
which are embedded into the symmetries group.

This general situation can however be mitigated in particular cases due to special extra structures a gauge natural theory may have.
For example, it has been discussed in details the case of spinors which are in fact a gauge natural theory for the spin group;
\cite{Kosmann}, \cite{Jadwisin}, \cite{Matteucci}. 
However,  due to the particular nature of the fields involved (precisely the frame acting as a spin frame and soldering form) one can define in such theories a sort of infinitesimal natural lift. 
In other words one is not able to lift arbitrary spacetime diffeomorphisms as in natural theories, but nevertheless can define a lift of one parameter subgroups of them. Moreover, such a lift commutes with the Lie algebra structure of the corresponding infinitesimal generators (i.e.~it preserves commutators) only when restricted to flows of isometries (i.e.~Killing vectors).
Accordingly, in special spacetimes (e.g.~on Minkowski) one can recover an action of flows of isometries on spinors and define a Lie derivative with respect to vector fields on the spacetime manifold. In this way one recovers completely the situation of special relativity when spinors on Minkowski space transform with respect to (infinitesimal) isometries.

Another example in which fields are gauge fields but one is able to define an action for general spacetime diffeomorphisms
is some formulation of Maxwell electromagnetism  as a natural theory; see \cite{ferraris}. 
 
In Section 2 we shall fix notation and introduce the natural structure for conformal gravity.
In Section 3 we shall obtain conservation laws induced by spacetime vector fields.
In Section 4 we shall discuss the physical relation between gauge natural and natural theories.

\section{Notation and naturality}

A field theory is defined on a {\it configuration bundle} $\calC$ with fibered coordinates $(x^\mu, y^i)$ where $x^\mu$ represent the position in a spacetime manifold $M$ of dimension $m$ (which is assumed to be paracompact, orientable and connected, and, for us, $m=4$). The coordinates $y^i$ represent instead fields. A configuration of the theory is a (global) section of  the configuration bundle, which is locally represented by providing field coordinates as a function of spacetime coordinates, namely by giving $y^i(x)$.

In gauge natural theories one starts from a suitable principal bundle $\calP$, which is called the {\it structure bundle},
and defines the configuration bundle as an associated bundle to $\calP$. 
In this way any (equivariant) automorphism on $\calP$ canonically induces an automorphism on $\calC$. Hence the configuration bundle $\calC$ comes with a selected 
subgoup of transformations  $\Aut(\calP)\subset \Aut(\calC)$ which will be requested to preserve the dynamics defined for the theory.
Such transformations are called {\it generalized gauge transformations}.
The automorphisms of $\calP$ canonically represented on the configuration bundle $\calC$ in turn define transformations on sections of $\calC$, i.e.~on fields.

The group of vertical automorphisms on $\calP$ will be hereafter denoted by $\Aut_V(\calP)\subset \Aut(\calP)$. 
It also induces automorphisms on the configuration bundle since $\Aut_V(\calP)\subset \Aut(\calP)\subset \Aut(\calC)$ and also induces transformations of fields which are called {\it pure gauge transformations}.
Finally, one can project generalized gauge transformations onto spacetime diffeomorphisms $\Diff(M)$ by declaring equivalent two automorphisms which project onto the same spacetime map. Accordingly, one has the following short exact sequence of groups
\begin{equation}
1\arr \Aut_V(\calP)\arr \Aut(\calP)\arr \Diff(M)\arr 1
\label{WitneySequence}\end{equation}
which captures the structure of different kinds of symmetry groups.

 We refer to \cite{Book} for the general framework of gauge natural theories; see also \cite{Kolar}, \cite{Eck}. The framework will be hereafter briefly exhibited for conformal gravity which is also explained in more details in \cite{Marta}.

 Let us choose a principal bundle $(P, M, p, \R)$ for the additive group $\R$ and let us denote the fibered coordinates as $(x^\mu, l)$. 
 Since $P$ is principal its transition functions are in the form 
 \begin{equation}
 \begin{cases}
 x'^\mu=x'^\mu(x)\\
 l'=\om(x)+ l\\
 \end{cases}
\end{equation}
 which are also affine transformations. Thus the bundle $P$ is at the same time principal and affine. Since it is affine it allows global sections.
 Since it is principal and it has global sections, it is trivial. Accordingly, $P$ is necessarily trivial and necessarily $P=M\times \R$.

 Moreover,  for any manifold $M$ one can define the bundle of general linear frames $(L(M), M, \pi, \GL(m))$, namely the principal bundle of  bases 
 of tangent vectors to $M$. The frame bundle $L(M)$ is a $\GL(m)$-principal bundle.
 
 One can paste these two principal bundles together to define the structure bundle $\calP=L(M)\times_M P$ with the group $\GL(m)\times \R$.
The configuration bundle $\calC$  is associated to the structure bundle by means of the suitable action  
\begin{equation}
\la: \GL(m)\times \R\times L\arr L: (J_a^c, \om, g)\mapsto g'_{ab}=e^\om \bar J_a^c g_{cd} \bar J^d_b
\end{equation} 
where $L$ denotes the set of symmetric, non-degenerate, bilinear forms of Lorentian signature $\eta=(3,1)$ and $g_{ab}$ are coordinates on $L$.

The configuration bundle is thence defined as $\calC=(L(M)\times P)\times_\la L$. Points in it are orbits of the action $[e_a, p, g_{ab}]_\la$ and  fibered coordinates are $(x^\mu, g_{\mu\nu})$ where we set $g_{\mu\nu}= e^l e_\mu^a g_{ab} e^b_\nu$ and $e^a_\mu$ denotes the inverse of the frame matrices 
$e_a=e_a^\mu \del_\mu$.

An automorphism on the structure bundles $\calP$ is locally represented as
\begin{equation}
 \begin{cases}
x'^\mu=x'^\mu(x)\\
 l'= \al(x) + l\\
 e'^\mu_a = J^\mu_\nu(x) \> e_a^\nu\\
 \end{cases}
\label{GGT}\end{equation}
where $J^\mu_\nu$ denotes the Jacobian of the spacetime  transformation and $\al$ is by now a pointwise element of the group $\R$, i.e.~a real valued function. 
Any generalized gauge transformation (\ref{GGT}) acts on the configuration bundle $\calC$ as
\begin{equation}
 \begin{cases}
x'^\mu=x'^\mu(x)\\
 g'_{\mu\nu}= e^{\al(x)}  J_\mu^\rho g_{\rho\si} \bar J^\si_\nu\\
 \end{cases}
\label{GGTC}\end{equation}
Notice that until here the generator of pure conformal transformations $\al$ and the spacetime diffeomorphism $x'^\mu(x)$ are completely unrelated.
Accordingly, the base manifold $M$ and the principal bundle $P$ chosen on it are unrelated. 
On the contrary $L(M)$ is functorially built out of the spacetime manifold $M$ and it contains the same information encoded in $M$. 

Then we choose a dynamics for which any of such transformation (\ref{GGTC}) is a symmetry. 
The Lagrangian density  for conformal gravity  (assumed quadratic in the curvature) turns out to be necessarily proportional to the squared Weyl tensor  (see \cite{Marta})
\begin{equation}
L= \sqrt{g} W_{\al\be\ga\de}W^{\al\be\ga\de}= \sqrt{g}\(R_{\al\be\ga\de}R^{\al\be\ga\de}- 2R_{\al\be}R^{\al\be}+\frac{1}{3} R^2\)
\end{equation}
The dynamics will induce algorithmically  conservation laws as we shall review below.

Before dealing with conservation laws let us go back in details to symmetry generators.
Let us choose a right-invariant pointwise basis for vertical vectors on the structure bundle. Let us set
\begin{equation}
\rho^\mu_\nu:= e_a^\mu \Frac[\del/\del e_a^\nu]
\qquad\qquad
\rho:= \Frac[\del/\del l]
\end{equation}  
which are right invariant as one can easily check.
Then any symmetry generator on $\calP$ is in the form
\begin{equation}
\Xi=\xi^\mu(x) \del_\mu + \del_\al \xi^\be(x) \rho_\be^\al + \ze(x) \rho
\label{SG}\end{equation}
Again up to now the component $\ze$ is unrelated to the components $\xi^\mu$ on spacetime.
Such a generator induces the infinitesimal generator of symmetries on configuration bundle in the form
\begin{equation}
\Xi_\la= \xi^\mu \del_\mu + \(  -\del_\mu \xi^\al g_{\al\nu} - g_{\mu\be}\del_\nu \xi^\be +\ze g_{\mu\nu}\)\Frac[\del/\del g_{\mu\nu}]
\end{equation}

The components of the vector field (\ref{SG}) transform as
\begin{equation}
\begin{cases}
\xi'^\mu= J^\mu_\al\xi^\al\\
\ze'= \ze + \xi^\mu\del_\mu \al\\
\end{cases}
\end{equation} 
and, as one can check, we have
\begin{equation}
\del'_\mu \xi'^\mu= \del_\mu\xi^\mu + \xi^\rho \bar J^\si_\al \del_\rho J^\al_\si
=\del_\mu\xi^\mu + \xi^\rho \del_\rho \ln J 
\end{equation}
where $J$ denotes the determinant of the Jacobian matrix $J^\al_\mu$ chosen in an (oriented) atlas.
This in fact suggests that the trace $\del_\mu\xi^\mu$ transforms as $\ze$ provided that transition functions $\al$ on $P$ are chosen so that
$\al = \ln J$.

Accordingly, on any (oriented) manifold $M$ we can fix an atlas and use its transition functions $J^\al_\mu$ to define a cocycle valued in $\R$ by using the
transition functions $\ln J$. This defines a natural principal bundle $\hat P$ which can be used to define the gauge natural theory as above.
The principal bundle $\hat P$ being natural can be explicitly built out as an associated bundle to $L(M)$. In fact let us set the action
\begin{equation}
\rho: \GL(m)\times \R\arr \R: (J^\mu_\nu, \om)\mapsto \om'= \ln J + \om  
\end{equation} 
Let us define the associated bundle $\hat P= L(M)\times_\rho \R$ which by construction has fibered coordinates $(x^\mu, l)$ which transform as
\begin{equation}
\begin{cases}
x'^\mu=x'^\mu(x)\\
l'= \ln J + l\\
\end{cases}
\end{equation}
One can see that the element $\ln J$ of group $\R$ acts by (left) translations onto $l$ so that the bundle $\hat P$ is by construction principal with the group $\R$.

Since any $\R$-principal bundle $P$ is trivial, then they are all isomorphic to each other, then $P\simeq \hat P\simeq M\times \R$ and the construction of $\hat P$ is simply a way of constructing the most general $\R$-principal bundle on $M$.
 However, since the structure bundle $\hat P$ now is also a natural bundle one can lift any spacetime vector field $\xi$ to it to obtain
 \begin{equation}
 \hat \xi= \xi^\mu\del_\mu + \del_\al \xi^\be(x) \rho_\be^\al + \del_\mu \xi^\mu \rho
\label{ConformalLift}\end{equation}
which, in turn, induces a vector field on configuration bundle  $C$ as
\begin{equation}
 \hat \xi_\la= \xi^\mu\del_\mu + \(  -\del_\mu \xi^\al g_{\al\nu} - g_{\mu\be}\del_\nu \xi^\be +\del_\al\xi^\al g_{\mu\nu}\)\Frac[\del/\del g_{\mu\nu}]
\end{equation}
Accordingly, the configuration bundle $\calC$ turns out to be a natural bundle as well.

If one considers standard conformal transformations on $M=\R^m$ (see \cite{Conformal}) their infinitesimal generators
$\xi= \xi^\mu\del_\mu\in \Vec(M)$ do in fact induces natural conformal transformations on $\calC$ generated by
$\hat \xi= \xi^\mu\del_\mu\in \Vec(\calC)$, i.e.~they satisfy the condition
\begin{equation}
\del_\al\xi^\al g_{\mu\nu}=  \del_\mu \xi^\al g_{\al\nu} + g_{\mu\be}\del_\nu \xi^\be 
\end{equation}
However, this condition makes a global sense only when the bundle $\calC$ is trivial, e.g.~when $M$ is parallelizable.
In general this definition would be coordinate dependent and local; the definition (\ref{ConformalLift})
is the only global, general prescription to let spacetime diffeomorphisms act on fields.

By standard techniques one can also show that any (torsionless) principal connection $\om=dx^\mu \otimes \(\del_\mu -\Ga^\al_{\be\mu}(x) \rho^\be_\al\)$ on $L(M)$
induces a connection on $\hat P$ which is given by
\begin{equation}
\te= dx^\mu\otimes \(\del_\mu - \Ga_\mu \rho\)
\end{equation}
where we set $\Ga_\mu=\Ga^\al_{\al\mu}$.

Given any symmetry generator (\ref{SG}) that can be split into the natural lift $\hat \xi$  of a spacetime vector field $\xi$ and a vertical field (which is a generator of pure conformal transformations), namely
\begin{equation}
\Xi= \(\xi^\mu\del_\mu + \del_\al \xi^\be(x) \rho_\be^\al + \del_\mu \xi^\mu \rho\) \oplus \( \ze - \del_\mu \xi^\mu\)\rho
=:\hat \xi \oplus \Xi_{(V)}
\end{equation}
The transformations generated by $\hat \xi$ will be called {\it natural conformal transformations}.
Thus any generalized conformal transformation can be split into a natural conformal transformation and a pure conformal transformation.

\section{Conservation laws}

In the previous Section we defined a natural framework for conformal theories on $\calC$
which needs to be compared with the gauge natural framework introduced in \cite{Marta}.  
In \cite{Marta} we already computed the conservation laws for the gauge natural  conformal theory.   As in any natural or gauge natural theory, one can show Noether currents admit superpotentials.
We found that pure conformal transformations do not contribute to superpotentials (for symmetry reasons); see \cite{tHooft}.

On the other hand, spacetime symmetries do define a superpotential
\begin{equation}
\begin{aligned}
\calU= \frac{1}{3}\Big\{& \left( 12 \na^{[\la} R^{\mu]}{}_\ep - 2\na^{[\la} R \de^{\mu]}_\ep \right) \xi^\ep +\\
+& 2\ \left( R g^{\nu[\mu} \de^{\la]}_\ep + 6 R^{\nu[\la} \de^{\mu]}_\ep + 3 R_\ep{}^{\nu\la\mu}\right) \na_\nu \xi^\ep \Big\} d \si_{\la\mu}
\end{aligned}
\end{equation}
where $d \si_{\la\mu}$ is the local standard basis for $(m-2)$-forms induced by coordinates.

In the gauge natural framework the pure conformal and spacetime transformations are completely unrelated.
In the natural setting instead,  one considers conservation laws generated by spacetime vector fields which act on the configuration bundle through their natural lift. In the natural lift the pure conformal part is fixed as a function of the vector field, namely by setting $\ze=\del_\mu\xi^\mu$.
However, as mentioned above the superpotential is totally independent of $\ze$ so also in the natural case the superpotential is entirely associated to generators of spacetime diffeomorphism. In other words conformal gravity specifically is insensitive to the way the spacetime vector fields are lifted to the configuration bundle just because the superpotential does not depend on pure conformal transformations. 

This seems to be the core of the issue raised in \cite{Jackiw} to argue against the physical meaning of conformal transformations as gauge transformations.

Since the theory exists in two different formulations, natural and gauge natural, but neither field equations or conservations laws seem to be different in the two approaches, one should ask if it really matters to choose one approach or the other or, instead, the two approaches have to be considered as equivalent.

\section{Natural vs.~Gauge Natural Formalism}

Conformal gravity is an example of a theory which can be formulated both as a gauge natural and a natural field theory.
As a gauge natural theory, pure conformal transformations are transformations on fields which do not affect the spacetime position.
In the natural framework spacetime diffeomorphisms (general diffeomorphism not just Killing vectors) act on fields and they (all) generate conservation laws.

One could thence ask whether one could not restrict to the natural framework from the beginning. After all pure gauge transformations do not generate conservation laws and probably it is worth discussing whether the two theories are really physically different in the end.

We showed that on any spacetime $M$ the bundle $P$ over it is trivial. 
Then there is only one principal bundle with group $\R$ over it, 
namely the trivial bundle  $P=M\times \R$. Since the construction of $\hat P$ is canonical, the natural bundle $\hat P$ coincides in fact with the only choice one has, i.e.~the trivial bundle. As a consequence, one always  has a global Lorentzian metric $g_{\mu\nu}$ to represent sections of $\calC$, i.e.~conformal structures on spacetime.

Thus if a difference between natural and gauge natural framework exists it is encoded into the physical meaning of some global property.
In particular one is not free to define the physical state of a theory at will, but some constraint is imposed by  dynamics.  
In fact one can apply a sort of hole argument to this situation; see \cite{Norton}, \cite{Covariance}.

Let us first notice that one can define compact supported  pure conformal transformations (assuming fields to be smooth,  not necessarily analytic).
Then considering any initial value problem defined on a Cauchy surface $S\subset M$ one can always find a pure conformal transformation which is supported on a compact $D\subset M$ which is not intersecting $S$. Since pure conformal transformations are symmetries, they map solutions into solutions and one can easily find two solutions with the same initial conditions on $S$. 

If one wants to save determinism for the physical state of this theory, one is basically forced to assume that the physical state is not associated to a single section of the configuration bundle (since we showed that for them there is no determinism since Cauchy theorem does not hold true) but rather classes of sections. In particular, in order to save determinism for the physical state one is forced to assume that configurations differing by a compactly supported pure conformal transformation do represents the same physical state. The dynamics in that case passes to the quotient and becomes deterministic.

(By the way, there is some freedom in defining precisely the physical state, since configuration differing by compactly supported symmetries have to represent the same state, but one may have some freedom to define bigger classes up to declaring that configurations differing by any symmetry represent the same state.)

This argument goes along the same lines used in electromagnetism to argue that gauge transformations do not affect the physical state. Then a family of local potentials $A_\mu$ which differ by gauge transformations on the overlaps defines a legitimate configuration for the electromagnetic field.
The, e.g.~the Dirac monopole, is a legitimate configuration of the electromagnetic field even if it is not defined on a trivial structure bundle.
In order to account for these configurations electromagnetism is a gauge natural theory and not a natural theory (even when the structure bundle does have a natural structure, as well; see \cite{ferraris}).

Thus we assume that in conformal gravity configurations differing by a (compactly supported) pure conformal transformations define the same physical state. The next issue is whether there exist metrics which are conformal without being diffeomorphic.
This is also easy to answer.
One can easily show that the scalar curvature $\tilde R$ of a conformal metric $\tilde g= \phi(x) \cdot g$ is related to the scalar curvature $R$ of the original metric $g$ as
\begin{equation}
\phi \tilde R= R-\frac{m-1}{\phi}\Dal \phi - \frac{(m-1)(m-6)}{4\phi^2} \na_\ep \phi \na^\ep \phi
\end{equation}
Accordingly, one can easily find a conformal factor $\phi$ for which $R$ and $\tilde R$ are different.
Since the scalar curvature is a diffeomorphism invariant, this shows that $\tilde g$ and $g$ are two configurations which are conformally equivalent though not diffeomorphic equivalent.
(One can easily use a standard partition of unity argument to find a compactly supported conformal factor, if needed.) 

The two metrics $\tilde g$ and $g$ represent two different physical states in the natural framework, while they represent the same physical state in a conformal theory.

Then in view of this result gauge natural conformal gravity has less states than natural conformal gravity and the two theories are, in this sense, not equivalent.
The decision between a natural and gauge natural formalism is dictated by the dynamics.
If the dynamics is preserved only by spacetime diffeomorphisms naturally acting on fields then the theory is natural and not gauge natural.

That is precisely why standard (purely metric) Hilbert-Einstein general relativity  is a natural theory and not a gauge natural theory for the structure bundle $L(M)$.
In fact the dynamics is covariant with respect to the group $\Diff(M)$ not for the group $\Aut(L(M))$.

For the same reason, in the tetrad formulation of standard GR the dynamics is covariant with respect to all gauge transformations of the Lorentz sub-bundle of $L(M)$ related to orthonormal frames. 
Accordingly, one is forced to resort to a gauge natural formalism with the Lorentz group as a gauge group; see \cite{Book}. 

In the same way, in conformal gravity the dynamics is covariant with respect to {\it any}  conformal transformation, not just the ones obtained by lifting spacetime diffeomorphims. 
Consequently, one should  have no choice than to regard conformal gravity as a gauge natural theory since its symmetry group includes all pure conformal transformations.

\section{Conclusions}

We showed that in conformal gravity the choice of whether regarding conformal transformations as spacetime transformations or gauge transformations is not really a matter of tastes. Since the dynamics is covariant with respect to general pure conformal transformations, then one is forced to 
regard conformal transformations as gauge transformations. 
Beside that we showed that if one overlooked the gauge nature of conformal transformations that will end up in a theory which is inequivalent to the correct one.
In particular, one distinguishes among configurations which are instead to be considered equivalent and Cauchy theorem fails to hold true for field equations.

Hence, unarguably, conformal gravity is a gauge natural theory. 
However, it also supports a natural lift which, unlike other gauge theories as, e.g., electromagnetism, one can define an action of diffeomorphisms
on fields. In other words the sequence (\ref{WitneySequence}) does in fact {\it canonically split}.

In particular, the action of diffeomorphisms allows to find a precise relation between results obtained for conformal transformations defined as pure field transformations and the corresponding results for conformal transformations defined as angle-preserving  maps on spacetime.
The two approaches have a perfect correspondence exactly in view of the natural structure. 
 
Future investigations should be devoted to extend this argument to more general conformal theories, other than conformal gravity,
especially to Minkowskian theories in which symmetries are extended to the conformal transformations group.

\section*{Acknowledgements}

We wish to thank A.Diaferio, N.Fornengo, C.Germani and So-Young Pi for useful discussions and comments. We also acknowledge the contribution of INFN (Iniziativa Specifica QGSKY), the local research project {\it  Metodi Geometrici in Fisica Matematica e Applicazioni (2014)} of Dipartimento di Matematica of University of Torino (Italy). This paper is also supported by INdAM-GNFM. One of us (M.C.) acknowledges partial support from the INFN grant InDark and from the grant Progetti di Ateneo/CSP TO\underbar Call2\underbar 2012\underbar 011 `Marco Polo' of the University of Torino and  thanks M.Meineri for valuable suggestions and support.


\begin{thebibliography}{99}

\bibitem{Jackiw}{R. Jackiw, So-Young Pi, 
{\it Fake Conformal Symmetry in Conformal Cosmological Models},
Phys. Rev. D 91, 067501;  arXiv:1407.8545 [gr-qc]
}

\bibitem{Mannheim}{P.D. Mannheim, D.Kazanas,
{\it Exact Vacuum Solution to Conformal Weyl Gravity and Galactic Rotation Curves},
Astro- phys. J. 342 (1989), pp. 635--638. 
}


\bibitem{Marta}{M.Campigotto, L.Fatibene,
{\it Gauge Natural Formulation of Conformal Theory of Gravity}, 
Annals of Physics, Volume 354, 328 (2015);  arXiv:1404.0898 [gr-qc]
}

\bibitem{Book}{L. Fatibene and M. Francaviglia, 
{\it Natural and Gauge Natural Formalism for Classical Field Theories},  
Kluver Academic Publisher, Dordrecht, 2003
}

\bibitem{200Lagrange}{M. Ferraris, M. Francaviglia, 
{\it The Lagrangian Approach to Conserved Quantities in General Relativity}, 
in: Mechanics, Analysis and Geometry: 200 Years after Lagrange, Editor: M. Francaviglia; Elsevier Science Publishers B.V., (1991), Amsterdam, 451--488
}

\bibitem{Augmented}{L. Fatibene, M. Ferraris,  M. Francaviglia,
{\it Augmented Variational Principles and Relative Conservation Laws in Classical Field Theory},
International Journal of Geometric Methods in Modern Physics 2 (June 2005), pp. 373--392; arXiv:math-ph/0411029.
}

\bibitem{Kosmann}{Y.Kosmann,  
{\it D\'eriv\'ees de Lie des spineurs},
Ann. di Matematica Pura et Appl. 91 317--395 (1972).
}

\bibitem{Jadwisin}{L. Fatibene, M. Francaviglia,
{\it Deformations of spin structures and gravity},
In: Gauge theories of gravitation (Jadwisin, 1997). Acta Phys. Polon. B 29 (1998), no. 4, 915--928.
}


\bibitem{Matteucci}{M.Godina, P.Matteucci,
{\it The Lie derivative of spinor fields: theory and applications}, 
Int. J. Geom. Methods Mod. Phys. 2 (2005) 159-188; arXiv:math/0504366 [math.DG]
}


\bibitem{ferraris}{M. Ferraris, M. Francaviglia, C. Reina,
{\it Sur les fibr\'es d'objets g\'eom\'etriques et leurs applications physiques},
 Ann. Inst. H. Poincar\'e, 38 (1983), 371-383.
}



\bibitem{Kolar}{I. Kolar, P. W. Michor, J. Slovak, 
{\it Natural Operations in Differential Geometry}, Springer- Verlag, New York (1993)
}

\bibitem{Eck}{D.J.~Eck,
{\it Gauge-natural bundles and generalized gauge theories},
Mem.\ Amer.\ Math.\ Soc.\ {\bf 33}(247) (1981)
}




\bibitem{Conformal}{P. Francesco, P. Mathieu, D. Senechal, 
{\it Conformal Field Theory}, 
Graduate Texts in Contemporary Physics, Springer (1997)
}

\bibitem{tHooft}{G. 't Hooft,
{\it Local Conformal Symmetry: the Missing Symmetry Component for Space and Time},
arXiv:1410.6675 [gr-qc], 2014
}


\bibitem{Norton}{J.D. Norton,
{\it General covariance, gauge theories and the Kretschmann objection},
in: Katherine Brading and Elena Castellani (eds.), Symmetries in Physics: Philosophical Reflections. Cambridge University Press. 110--123 (2003)}

\bibitem{Covariance}{L.Fatibene, M.Ferraris, G.Magnano,
{\it Constraining the Physical State by Symmetries},
(in preparation)
}









\end{thebibliography}
\end{document}